\documentclass[journal]{IEEEtran}
\usepackage{nopageno}

\usepackage{amsmath}
\usepackage{algorithmic}

\usepackage{amsmath}
\usepackage{mathrsfs}
\usepackage{amssymb}
\usepackage{float}
\usepackage{siunitx}
\usepackage{amsmath}
\usepackage{import}
\usepackage{mathtools}
\usepackage{pdfpages}
\usepackage{pgf}
\usepackage{pgfplots}
\usepackage{tikz}
\usepackage{pgfgantt}
\usepackage{relsize}
\usepackage{mathptmx} 
\usepackage{trfsigns} 
\usepackage{placeins}
\usepackage{graphicx, caption, subcaption}

\usepackage{todonotes}
\usetikzlibrary{positioning}
\pgfplotsset{compat=newest}
\pgfplotsset{plot coordinates/math parser=false}

\usepackage{subcaption}

\usepackage{fixltx2e}
\usepackage{xcolor}
\begin{document}

\title{Disassemblable Fieldwork CT Scanner Using a 3D-printed Calibration Phantom}

\author{Florian Schiffers, Thomas Bochynek, Andr\'e Aichert, Tobias W{\"u}rfl, Michael Rubenstein, Oliver Cossairt
	\thanks{F. Schiffers and T. Bochynek are with the Department of Computer Science, Northwestern University, IL, Evanston, 60201, USA}
	\thanks{M. Rubenstein and O. Cossairt are with the Department of Computer Science and Electrical and Computer Engineering, Northwestern University, IL, Evanston, 60201, USA}
	\thanks{A. Aichert is with Digital Technology Innovation, Siemens Healthineers, Erlangen, Germany}
	\thanks{T. Wuerfl is with the Chair of Pattern Recognition, FAU Erlangen, Germany.}
	\thanks{Corresponding author: florian.schiffers@northwestern.edu}

}

\maketitle

\begin{abstract}

The use of computed tomography (CT) imaging has become of increasing interest to academic areas outside of the field of medical imaging and industrial inspection, e.g.\, to biology and cultural heritage research.
The pecularities of these fields, however, sometimes require that objects need to be imaged on-site, e.g., \ in field-work conditions or in museum collections.
Under these circumstances, it is often not possible to use a commercial device and a custom solution is the only viable option.
In order to achieve high image quality under adverse conditions, reliable calibration and trajectory reproduction are usually key requirements for any custom CT scanning system.
Here, we introduce the construction of a low-cost disassemblable CT scanner that allows calibration even when trajectory reproduction is not possible due to the limitations imposed by the project conditions.
Using 3D-printed in-image calibration phantoms, we compute a projection matrix directly from each captured X-ray projection.
We describe our method in detail and show successful tomographic reconstructions of several specimen as proof of concept.
\end{abstract}


\IEEEpeerreviewmaketitle

\thispagestyle{empty}
\section{Introduction}

\IEEEPARstart{W}{hen} computed tomography (CT) was invented in 1971 it quickly became an essential tool for non-destructive testing and medical diagnosis.
The proliferation of CT hardware and software in recent decades has promoted widespread interest from the greater scientific community, finding new applications in a broad range of fields such as cultural heritage~\cite{Stromer2019,Morigi2010} and biology~\cite{Singh2019}.

While industrial CT applications often feature optimized and standardized imaging setups that allow high-throughput imaging, e.g.\ in a hospital setting or a production line, the scientific community operates under a fundamentally different set of constraints than medical and industrial applications.
For instance, biological and cultural heritage research often prohibits samples from being transported to the nearest CT facilities, requiring scanning on-site.

Under these circumstances existing commercial products are unsuitable, and a custom-built, portable system is necessary.
Of particular interest here is the case where the imaging system must be shipped between sessions, and repeated calibration is necessary.
As a potential solution to this problem, Aichert~\emph{et al.}~\cite{Aichert2018a} proposed an online geometry calibration method using in-image fiducial markers for flat-panel computed tomography based on projective invariants that can be easily realized with a 3D-printed metal bead phantom.
They demonstrate a simulation-based proof-of-concept for a calibration phantom designed for computing both extrinisic and intrinsic camera parameters from a single projection image.

This paper describes the construction of a novel and disassemblable cone-beam computed tomography (CBCT) scanner, which for the first time, applies the calibration method proposed in~\cite{Aichert2018a}.
We show reconstructions of biological specimen imaged on Barro Colorado Island, a remote, tropical island in Panama.
This work demonstrates that customized imaging and calibration setups allow for the acquisition of industry-standard quality images under demanding imaging environments.

\begin{figure}[t]%
\centering
	\begin{subfigure}[t]{0.48\columnwidth}
		\includegraphics[width=\textwidth]{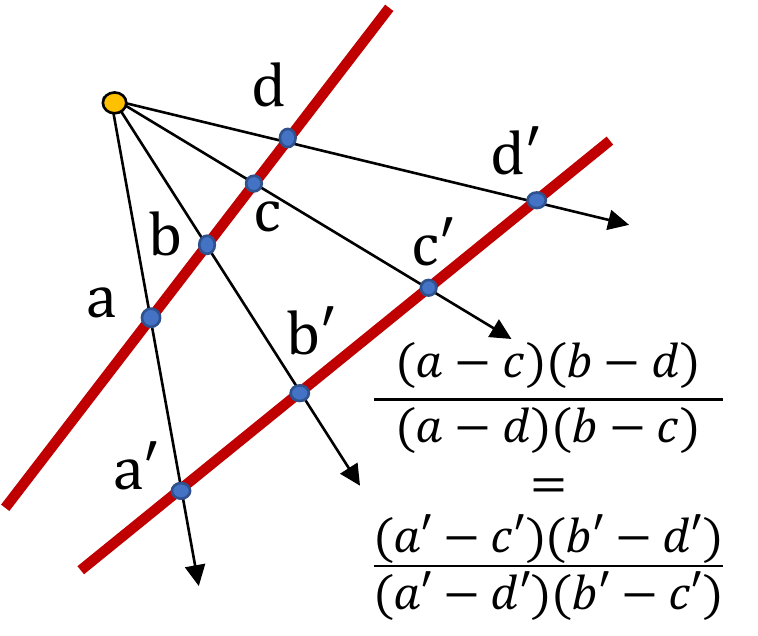}
		\caption{Cross-Ratio }
	\end{subfigure}
	\hfill
	\begin{subfigure}[t]{0.48\columnwidth}
		\includegraphics[width=0.95\textwidth]{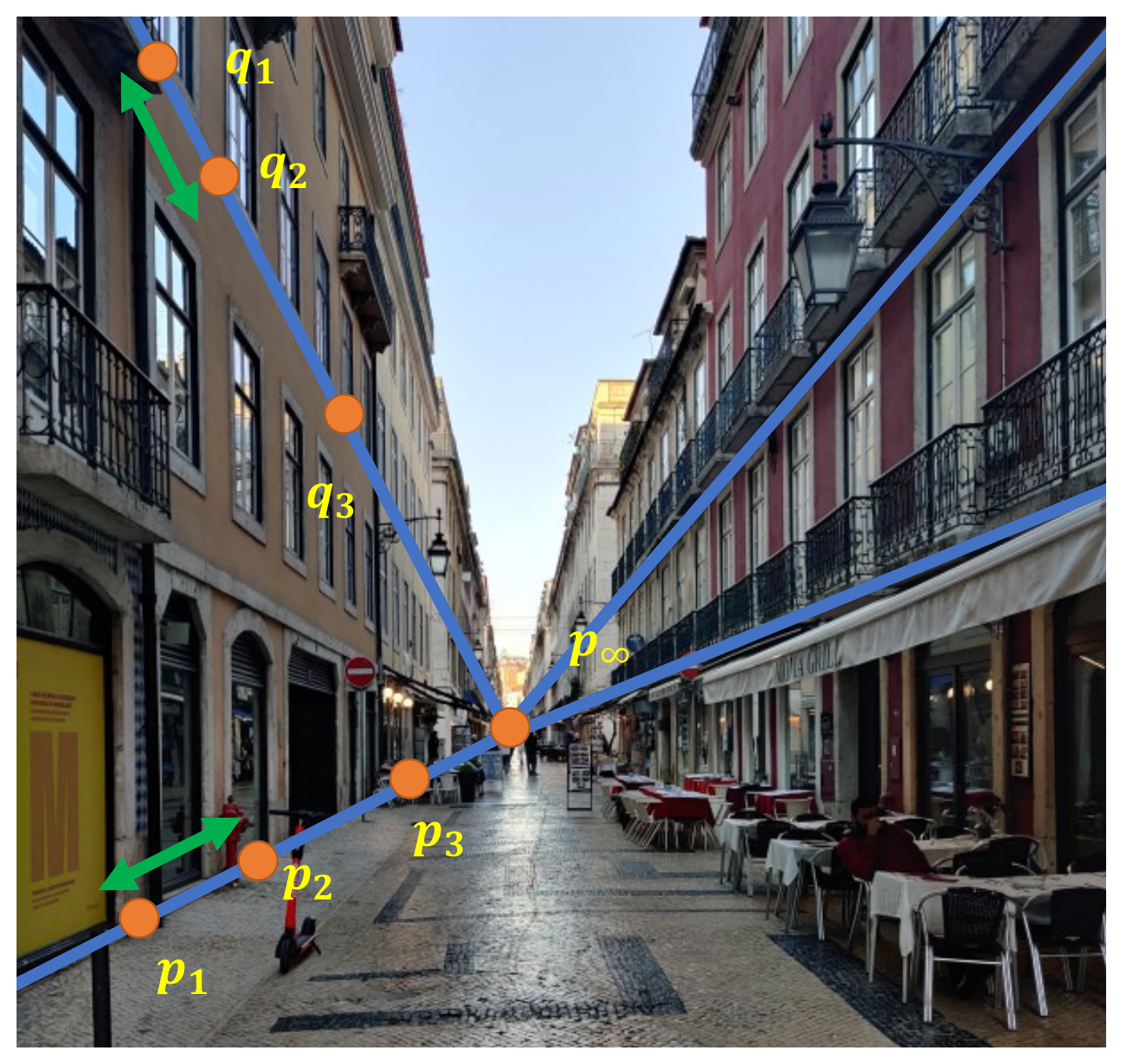}
		\caption{Cross-section to calculate relative distances}
	\end{subfigure}

	\caption{Schematic principle of the projective invariant. }
	\label{fig:cross_ratio_example}
	\vspace{-0.4cm}
\end{figure}



\section{Methods}

\subsection{Cone Beam Projection Model}
The most popular trajectory takes the shape of an arc of a circle.
Such trajectories depend on few parameters compared to the calibration of the geometry of each projection image~\cite{VonSmekal2004,Daly2008}.
Therefore it is appealing to assume an ideal smooth trajectory and calibrate only those parameters as opposed to estimating a geometry for each projection image independently.
This simplification, however, comes at the cost of no longer being able to compensate for deviations from this ideal smooth trajectory.
An alternative is to describe each projection matrix independently e.g.~\cite{VonSmekal2004}.
For those methods, the image formation of a conventional X-Ray source with a cone beam and flat-panel detector is typically modeled as a projection matrix.

In an industrial CT setup, the focal length of the system and the principal point, which is the intersection of the orthogonal projection of the source to the detector, are stable because the source and detector do not move during the acquisition.
Therefore the intrinsic camera parameters can be measured as fixed in the whole system, and only the extrinsic camera parameters need to be determined for each projection.

\begin{figure}[t]

	\begin{subfigure}[t]{0.48\columnwidth}
	\centering
		\includegraphics[width=\textwidth]{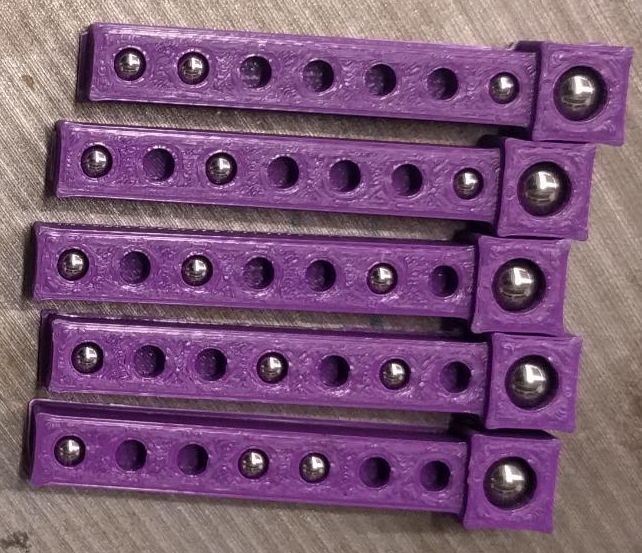}
		\caption{3D-printed markers with unique metal bead placement to achieve different cross-ratios.}
		\label{subfig:sticks}
	\end{subfigure}
	\hfill
	\begin{subfigure}[t]{0.48\columnwidth}
	\centering
		\includegraphics[width=0.8\textwidth]{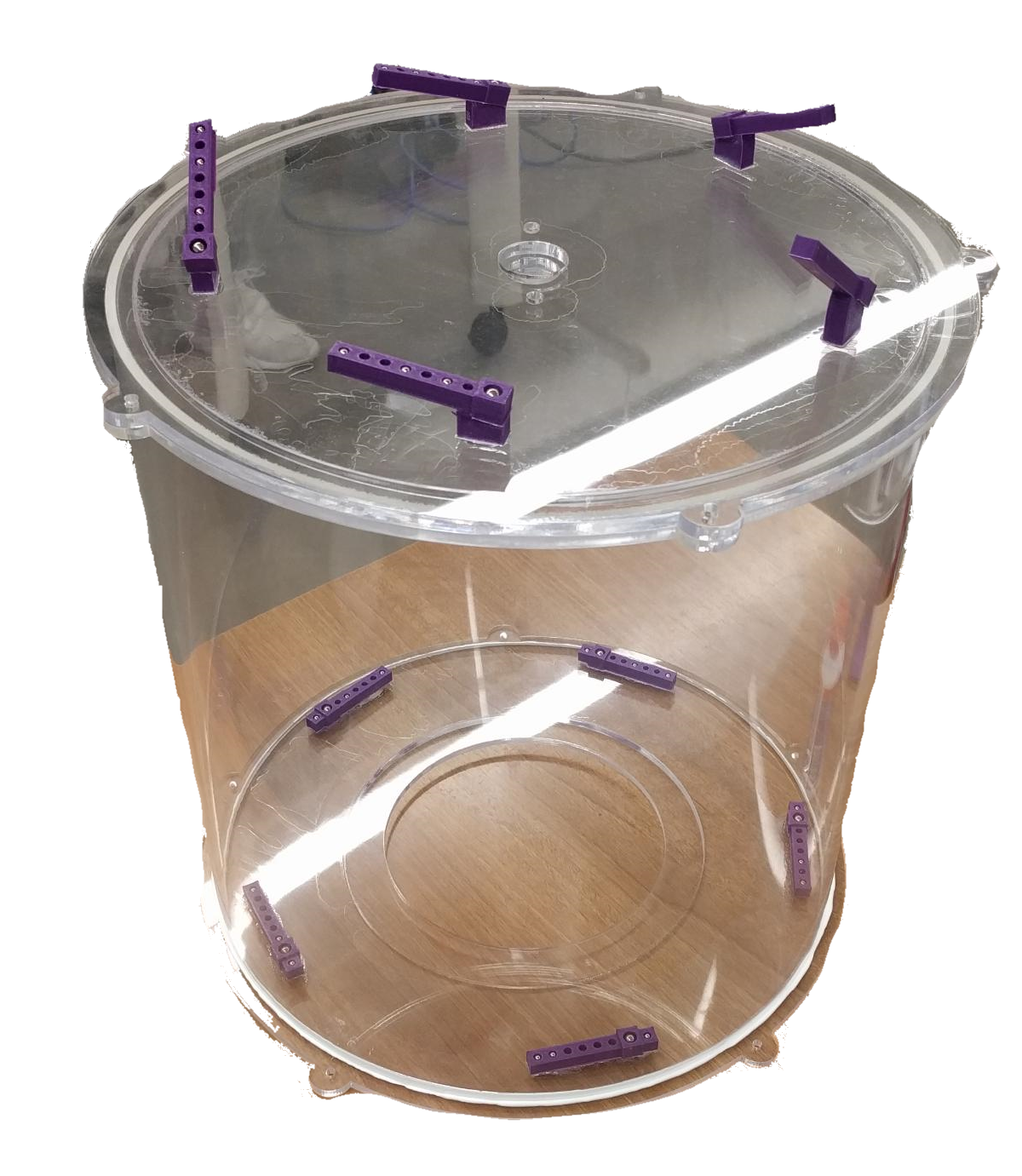}
		\caption{Sample container with calibration phantom consisting of ten calibration markers}
		\label{subfig:container}
	\end{subfigure}

	\caption{3D-printed calibration markers, in detail and assembly.}

\end{figure}

\subsection{Estimation of Geometric Parameters}

A significant challenge for tomographic calibration is to build a 3D-phantom with interest points that can be uniquely identified under a projective transformation in different projections. A pair of coordinates of the same 3D interest point in the projections are called corresponding points and the problem of computing them is commonly referred to as the correspondence problem.
Strobel~\emph{et al.}~\cite{Strobel2003} proposed the \emph{PDS-2} phantom where binary codes of small metal beads are encoded to solve the correspondence problem.
Choi~\emph{et al.}~\cite{Choi2014} solve this problem using the assumption that markers only move a small distance in successive projections.

These and similar methods usually acquire a dedicated calibration scan, since CT scanners are manufactured to consistently reproduce the same motion.


In contrast, Aichert~\emph{et al.}~\cite{Aichert2018a,Aichert2019} present a method relying on the cross-ratio to solve the correspondence problem. If the phantom is present in the acquisition the method can be used to calculate a projection matrix for each X-Ray projection, which removes the necessity of reproducible trajectories. This approach meets the requirement of robust reconstruction under demanding field conditions. Each phantom consists of multiple, short calibration elements of four uniquely spaced metal beads each.

These elements can be arranged arbitrarily within the projection space but in fixed relation to the sample (e.g.\,  by attaching them to a sample container).
The respective cross-ratio of the metal beads, depicted in Fig.~\ref{fig:cross_ratio_example}, is used as a scalar descriptor for each such calibration element.
It is defined by the distances $a, b, c, d$ $\in \mathbb{R}$  between the four metal beads in a linear arrangement.

\begin{equation}
\text{cr}(a,b,c,d) := \frac{(a-c)\cdot (b-d)}{(a-d) \cdot (b-c)} \,.
\end{equation}

Because this number is invariant to perspective projection, it can be used to identify corresponding points in different projections.
A RANSAC-based algorithm is used to estimate all geometric parameters from a single image.
For a detailed discussion of the mathematical methodology and how the cross-ratio is used for calibration, we refer to~\cite{Aichert2018a}.

\subsection{Phantom Design and Manufacture}
We fabricated ten calibration elements of six-centimeter length, each containing a unique arrangement (and thereby cross-section) of four metal beads.
To provide directionality, the diameter of the lower-most bead was twice that of the remaining beads (i.e., \ 3.2 mm and 1.6 mm, respectively).


We 3D-printed the base marker shape from Polylactic acid.
Metal ball bearings were press-fitted into predefined recesses.
Fig.~\ref{subfig:sticks} exemplary shows five of those calibration sticks. The assemblage of the ten sticks in 3D space is such that at least three are fully visible on the detector for any projection image of the trajectory.

We placed five of these calibration elements, each to both the top and bottom of the sample container, see Fig.~\ref{subfig:container}.
Equidistant arrangement of the elements ensured that there would be three markers identifiable for any projection image of the trajectory, removing marker overlap as a possible complicating factor.
To avoid overlap of markers and the sample specimen, we reserve the top and bottom of the projection for the calibration phantom.
Note that this effectively trades image information for a reliable and accurate online estimation of geometric parameters.



\begin{figure}[t]

	\begin{subfigure}[t]{0.48\columnwidth}
		\includegraphics[width=\textwidth]{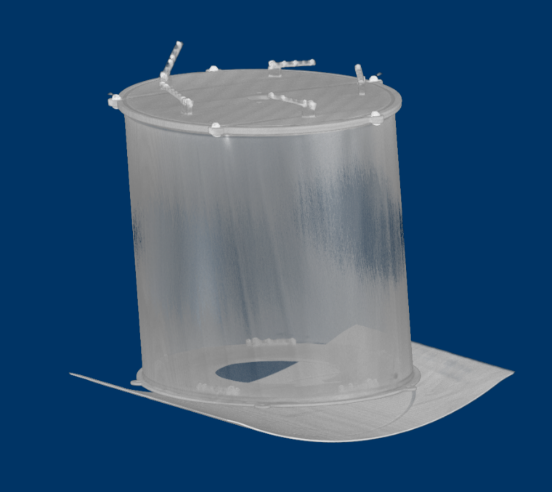}
		\caption{CT of 3D-printed calibration phantom with metal beads.}
		\label{subfig:ct_scan_phantom}
	\end{subfigure}
	\hfill
	\begin{subfigure}[t]{0.48\columnwidth}
		\includegraphics[width=\textwidth]{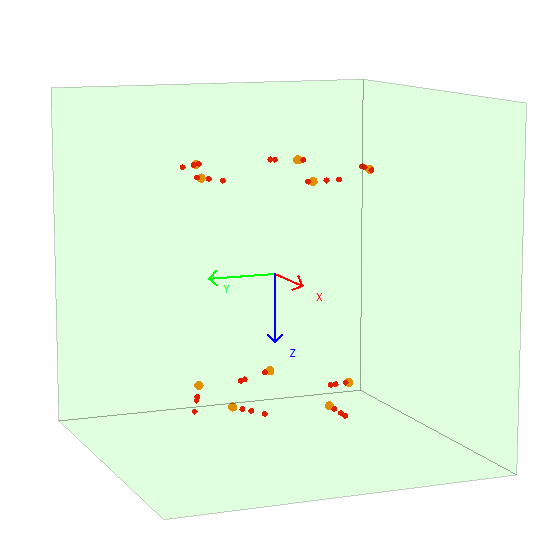}
		\caption{Extracted 3D positions of metal beads.}
		\label{subfig:ct_extracted_3d_positions}
	\end{subfigure}

	\caption{The 3D-printed calibration is imaged with a commercial CT scanner at a hospital. The 3D locations of the metal beads are extracted from DICOM-series. }

\end{figure}

\subsection{Hospital CT-scan of Calibration Phantom}
The algorithm proposed in \cite{Aichert2018a} requires knowledge of the exact 3D positions of all beads in the phantom.
While it is known from computer vision that these positions could be inferred algorithmically, we instead acquired a reference scan of the whole calibration phantom with a commercial CT scanner in the Northwestern Memorial Hospital (see Fig.~\ref{subfig:ct_scan_phantom}).

The spheres are identified from the DICOM series via the application of the connected components method. Subsequently we calculate the center of mass of each sphere as shown in Fig.~\ref{subfig:ct_extracted_3d_positions}.
Note, that the marker accuracy is given not by the accuracy of the 3D-printing process, but by the identification of the circle centers.

\subsection{Experimental Setup}

\begin{figure}[t]
	\includegraphics[width=\columnwidth]{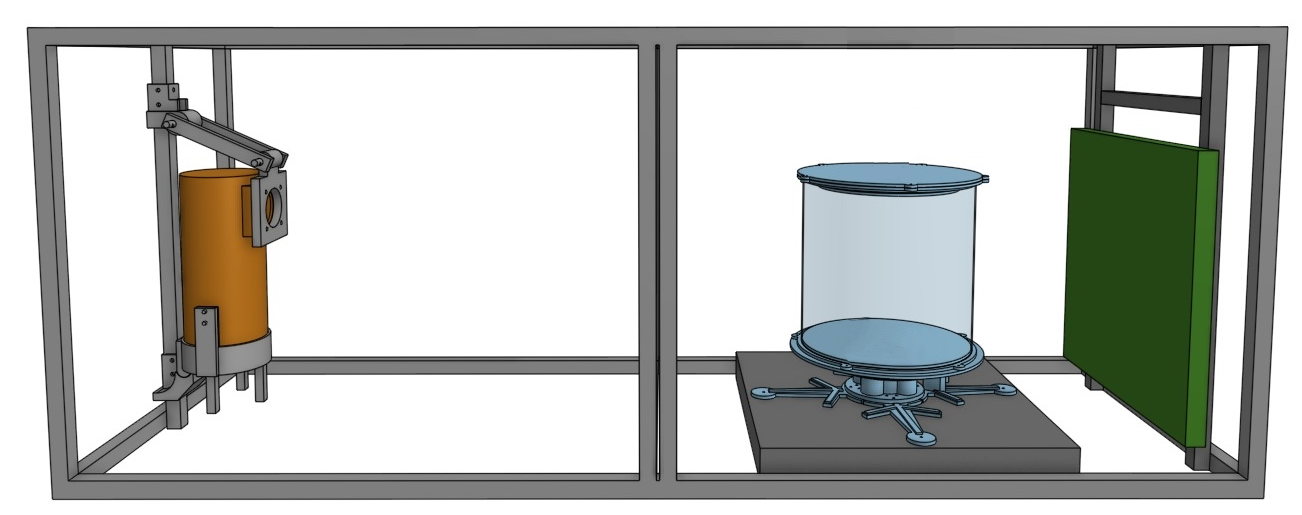}
	\caption{Computer model of the scanner setup.
	Mount for the X-Ray source (left) and rotation stage (center right) are fabricated from 3D-printed components. See text for details on the used components.
	\label{fig:cad_design}
	}
\end{figure}

The CT scanner setup (see Fig.~\ref{fig:cad_design}) is composed of four main components:
The imaging container with the calibration phantom is located inside an 80/20 T-slot aluminum-frame cabinet, which is shielded by fused lead/steel plates for radiation protection.
We used a \textit{Imaging M-113T X-Ray tube} and a \textit{Varex Imaging XRD 4343 Digital X-Ray detector} for radiation generation and detection.
A rotation stage, driven by a \textit{1000:1 Micro Metal Gearmotor HP 6V}, rotated the imaging container around its axis.
Rotation stage, X-Ray tube, and X-Ray detector were controlled and synchronized via an \textit{Arduino Uno microprocessor}.

A single 3D scan consisted of approximately 450 exposures of the rotating imaging container at a rate of 4.5 exposures per second.
A single exposure consisted of a 10.6 ms X-Ray burst at 40 keV energy and 125 mA current supplied by a \textit{Spellman PMX 5kW X-Ray generator}.

\section{Reconstruction pipeline}

We implemented the reconstruction pipeline in Python and embedded our pipeline into the open-source framework of \mbox{pyCONRAD} \textit{(https://git5.cs.fau.de/PyConrad/pyCONRAD)} which natively supports arbitrary trajectories via projection matrices~\cite{Maier2013}.
Since pyCONRAD has the full GPU functionality of CONRAD, fast reconstruction with the GPU-accelerated back-projector is possible.
For reconstruction, we use the Feldkamp algorithm~\cite{Feldkamp1984}.

\subsection{Image Pre-Processing}
Raw images are normalized using the following normalization equation:
\begin{equation}
f = \frac{I - I_{Dark}}{I_0 - I_{Dark}}
\end{equation}
where $I$ denotes the measured intensity with the object, $I_0$ the reference intensity without object, and $I_{Dark}$ the measured dark-image.

Our detector has approximately 100 permanent defect pixels which either report full or zero intensity in every exposure.
If not accounted for, these pixels will result in strong ring-artifacts in the reconstructions.
We hence detect those pixels with a prior calibration measurement and correct them using OpenCV's inpainting method.

\subsection{High resolution volume reconstruction}
Our detector allows for a resolution of up to $3000 \times 3000$ pixels, generating large volumes of data of both raw data and reconstructions.
For example, a reconstructed volume of $2000^3$ with float precision requires approximately $32$ GB of storage space.
PyCONRAD does not natively support reconstruction volumes of this size because it is limited by the GPU memory.
Reconstruction of large high-resolution volumes was achieved by successively reconstructing and storing smaller volume subsets.

\subsection{Results}
\begin{figure}%

	\begin{subfigure}[b]{0.48\columnwidth}
	\includegraphics[width=\textwidth]{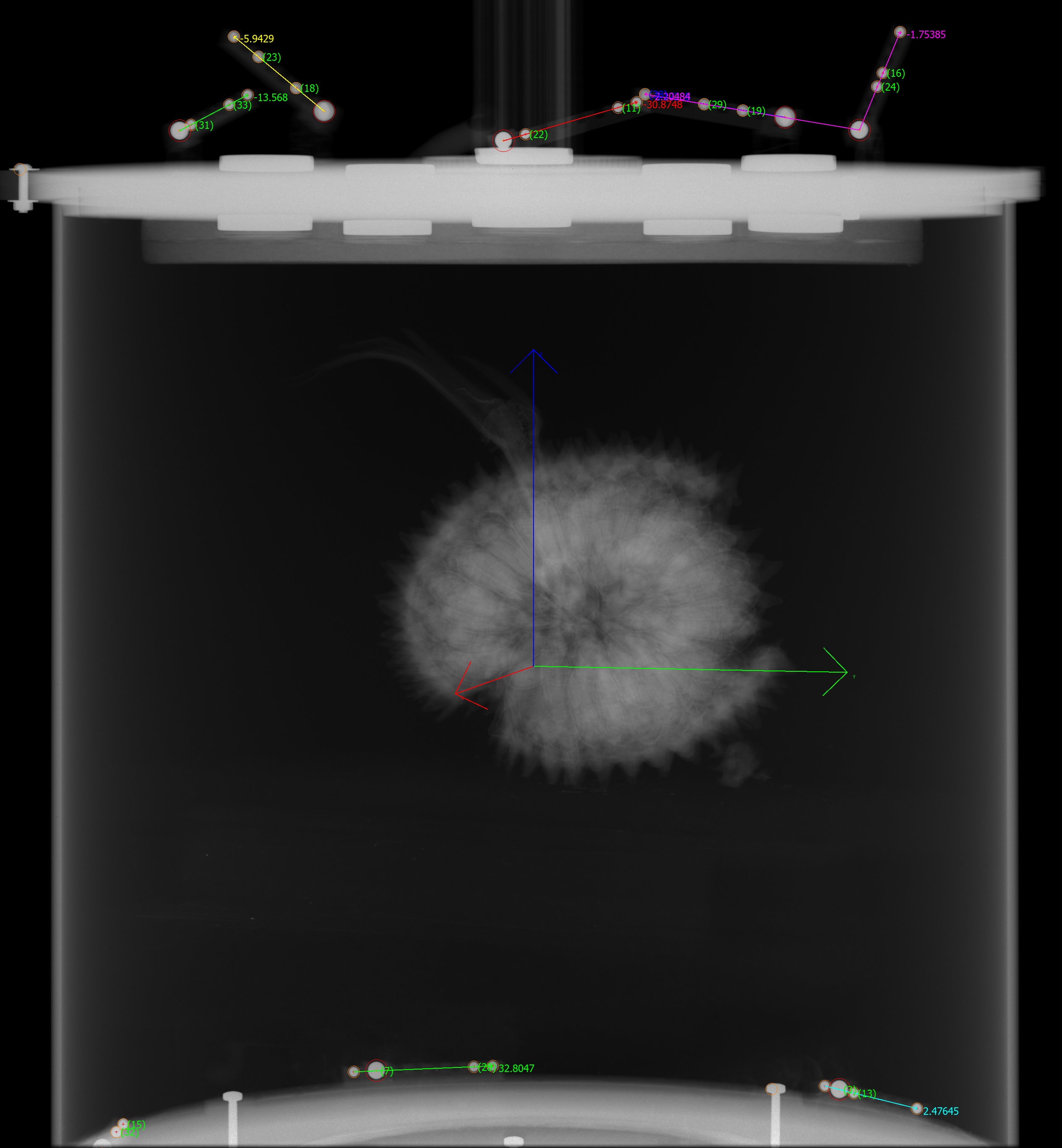}
	\caption{Projection image 1}
	\end{subfigure}
	\hfill
	\begin{subfigure}[b]{0.48\columnwidth}
	\includegraphics[width=\textwidth]{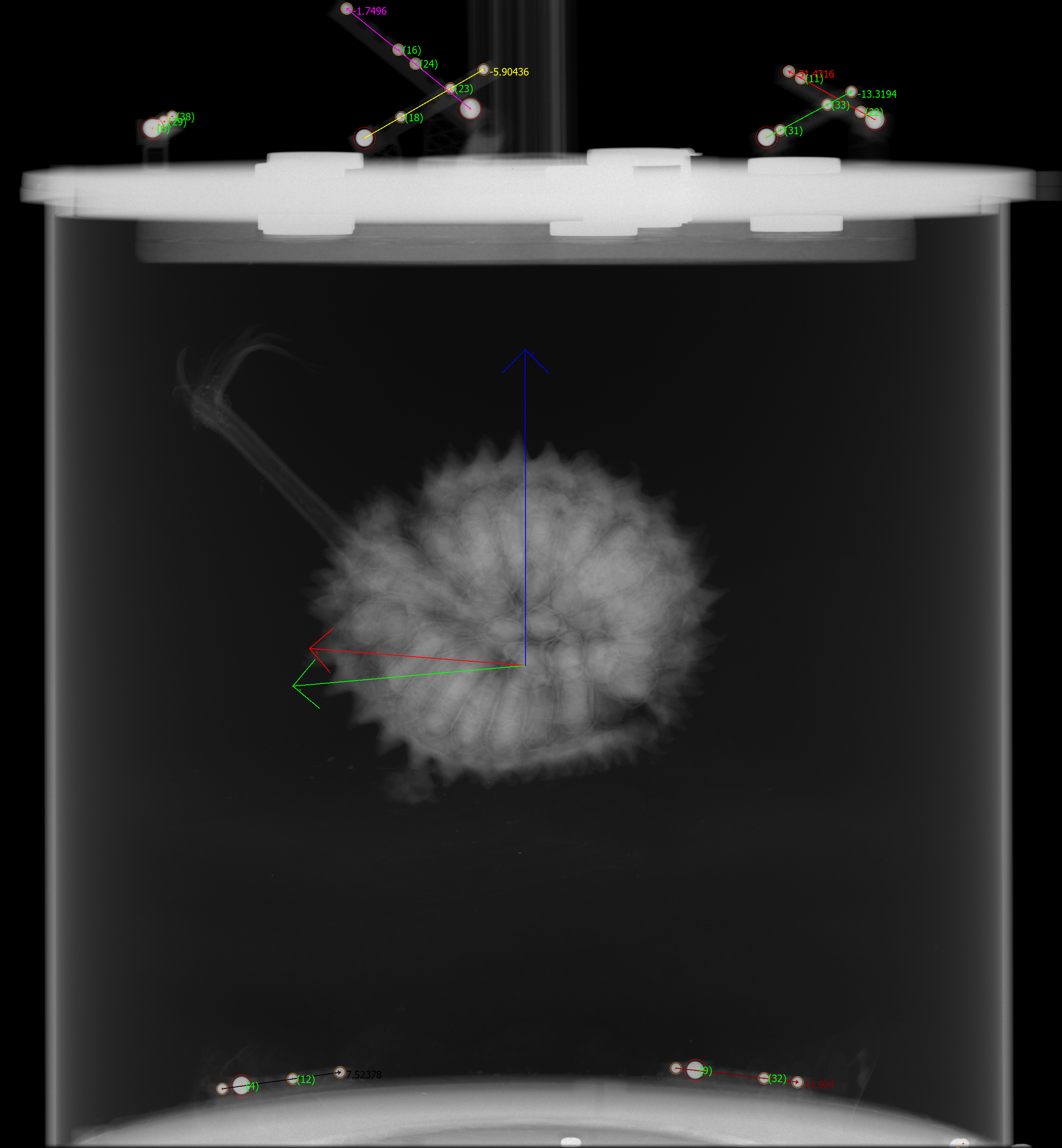}
	\caption{Projection image 2}
	\end{subfigure}

	\begin{subfigure}[b]{0.48\columnwidth}
	\includegraphics[width=\textwidth]{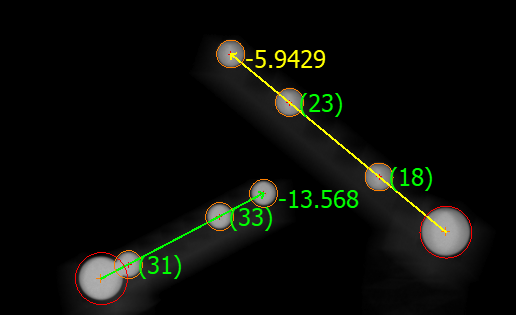}
	\caption{Close-up showing reprojection error 1}
	\end{subfigure}
	\hfill
	\begin{subfigure}[b]{0.48\columnwidth}
	\includegraphics[width=\textwidth]{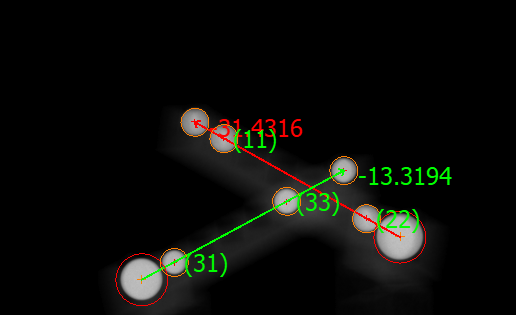}
	\caption{Close-up showing reprojection error 2}
	\end{subfigure}

\caption{(a,b) Projection images with biological specimen (\textit{Annona purpurea} fruit) and calibration markers.
Detected calibration elements and the derived local coordinate system are shown in the center of each image. (c,d)~show a zoomed version for better visualization of the detected beads' reprojection.}

\label{fig:calibration_fruit_example}

\end{figure}

\begin{figure}[!ht]%

\begin{subfigure}[t]{\columnwidth}
        \centering
	\includegraphics[width=0.4925\textwidth]{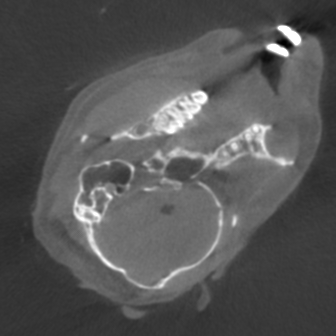}
\hfill
	\includegraphics[width=0.4925\textwidth]{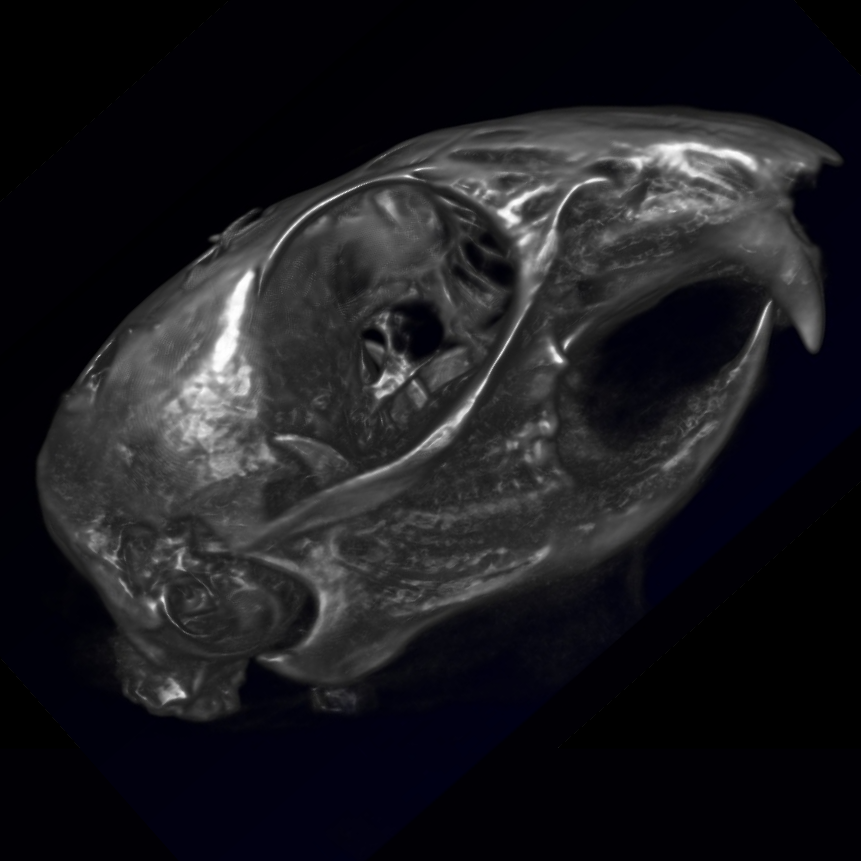}
\caption{\textit{Tamiasciurus sp.\:} squirrel skull}
\vspace{1mm}
\end{subfigure}
\vspace{1mm}
\begin{subfigure}[t]{\columnwidth}
        \centering
	\includegraphics[width=0.4925\textwidth]{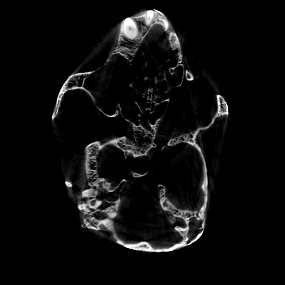}
\hfill
	\includegraphics[width=0.4925\textwidth]{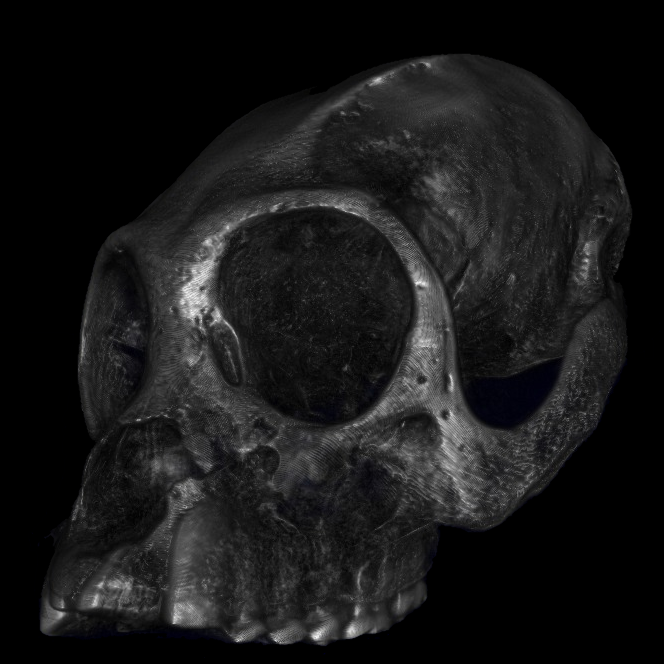}
\caption{\textit{Alouatta palliata:} howler monkey skull }
\end{subfigure}
\vspace{1mm}
\begin{subfigure}[t]{\columnwidth}
        \centering
	\includegraphics[width=0.4925\textwidth]{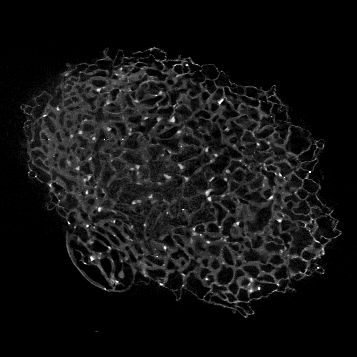}
\hfill
	\includegraphics[width=0.4925\textwidth]{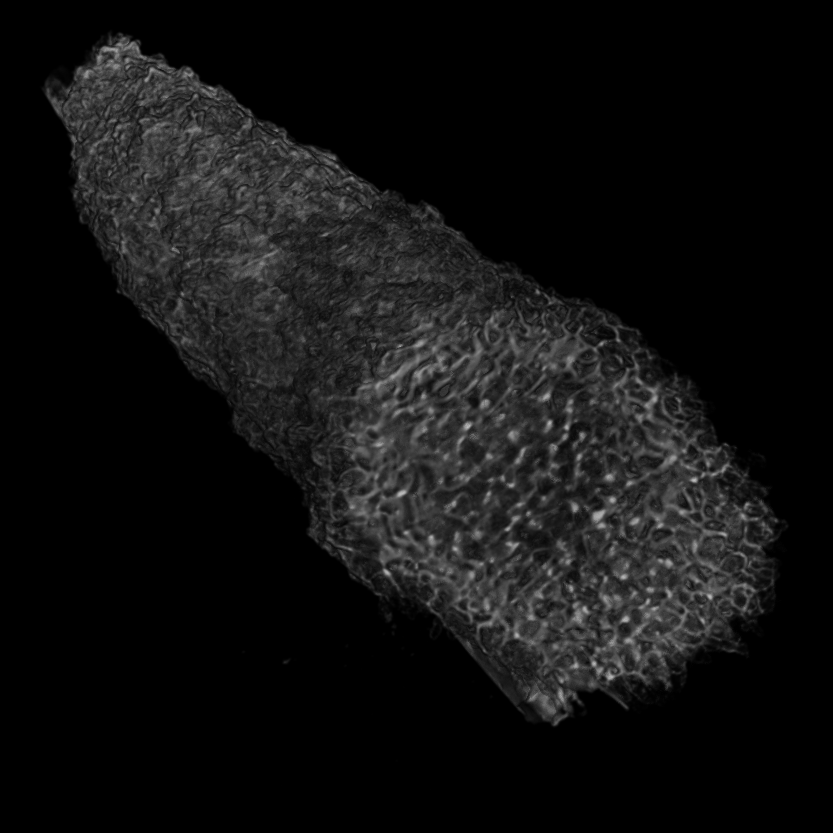}
\caption{\textit{Nausititermes sp.\:} termite nest}
\end{subfigure}
\begin{subfigure}[t]{\columnwidth}
        \centering
	\includegraphics[width=0.4925\textwidth]{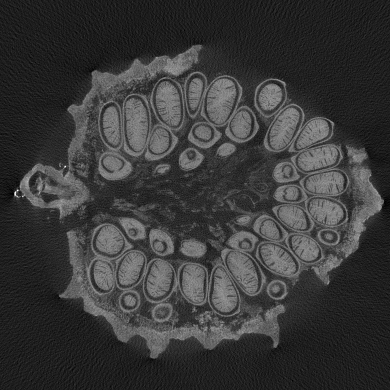}
\hfill
	\includegraphics[width=0.4925\textwidth]{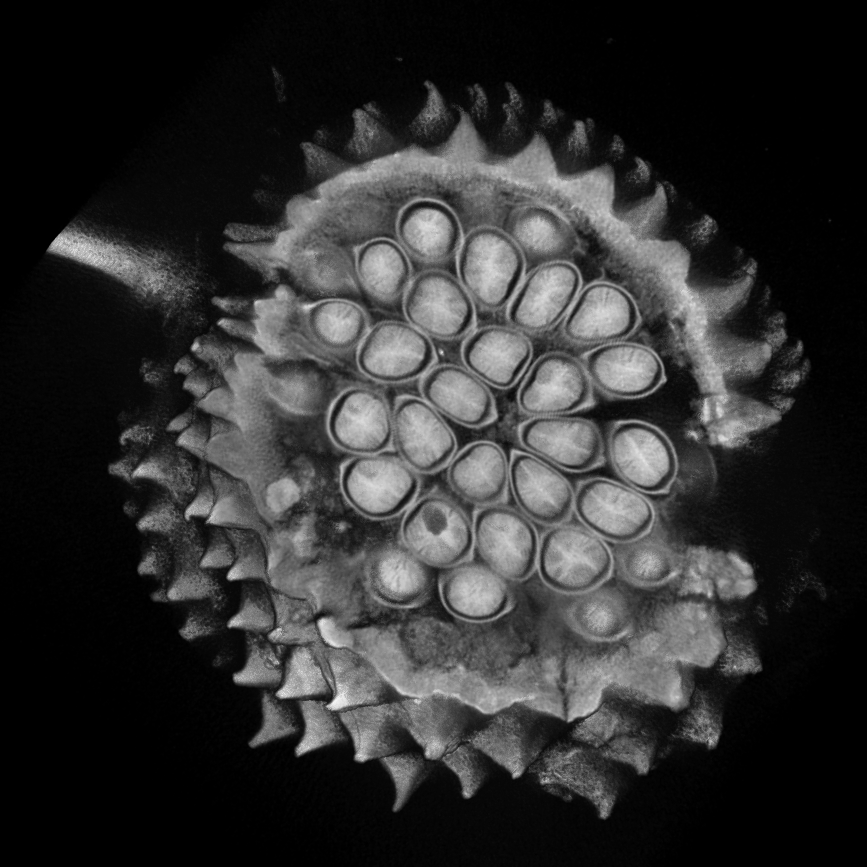}
\caption{\textit{Alouatta palliata:} tropical fruit}
\end{subfigure}


\caption{Example reconstruction from our tomographic scanner. Left: Tomographic slice; right: volume rendering.}
\label{fig:example_reconstructions}
\end{figure}

\begin{figure}%

  \begin{subfigure}[b]{0.48\columnwidth}
	\includegraphics[width=\textwidth]{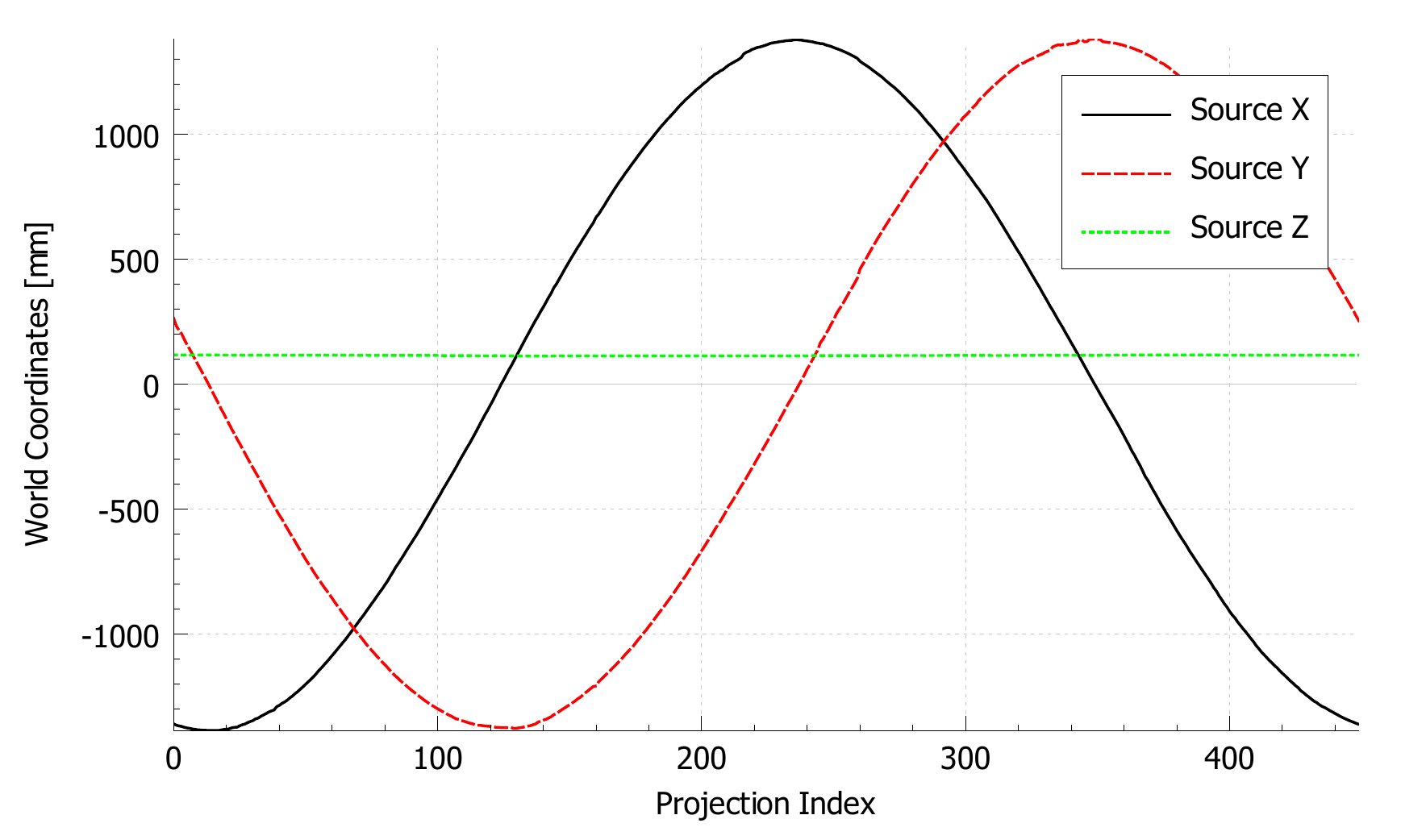}
	\caption{Calculated source position}
\end{subfigure}
\hfill
\begin{subfigure}[b]{0.48\columnwidth}
	\includegraphics[width=\textwidth]{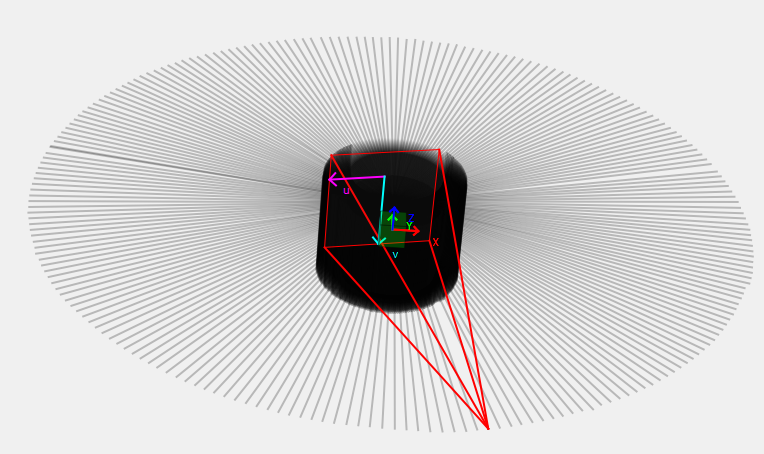}
	\caption{Calculated trajectory}
\end{subfigure}

\caption{Visualization of the trajectory calculated online during a tomographic scan.
The image reveals that the source positions do not form perfect sinusoids, and the angular spacing is not equidistant.
While this would be a challenge to a one-time calibrated setup, the online calibration we employ compensates for these inaccuracies.
}

\label{fig:calibration_example}

\end{figure}

Figure~\ref{fig:calibration_fruit_example} shows an example of an online-calibration with bead detection, feature matching using the cross-ratios, and successful geometric calibration for two different views. The complete trajectory was calibrated and is visualized in Fig.~\ref{fig:calibration_example}. Example tomographic reconstructions and renderings for four specimens found on-site at the Barro Colorado Island are shown in Fig.~\ref{fig:example_reconstructions}.

\section{Discussion and Conclusion}
We constructed a 3D CT scanner and online calibration phantom from commonly available materials using 3D printing as a key manufacturing method.
While the methods we employ in the construction process might not meet the high-accuracy manufacturing requirements applied to medical devices, we still obtain high-quality 3D reconstructions.
In contrast to one-time calibrated and commercial scanners, our setup is resilient to adverse conditions, disassemblable for transport to remote locations, and was successfully operated for field-work in a tropical environment.
We intend that this paper can serve as a guideline for interested parties to develop their own applications.


Future research directions include reducing the amount of detector space required for online calibration, employing bundle-adjustment to lift the requirement of a prior CT-scan of the calibration phantom,
using our methodology on a theoretically complete trajectory like a helix,
additional artifact reduction techniques, such as beam-hardening correction, as well as the use of image-based calibration correction methods.


\bibliographystyle{ieeetr}

\end{document}